\documentclass[aps,pra,twocolumn,showpacs,groupedaddress,amsmath,amssymb]{revtex4}

\usepackage{bm}
\usepackage{graphicx}
\usepackage{amsmath}

\newcommand{\lsup}[2]{\mbox{$^{#1}{#2}$}}
\newcommand{\onemu}{\mbox{$\mspace{1.0mu}$}}
\makeatletter
\def\gsim{\compoundrel>\over\sim}

\def\compoundrel#1\over#2{\mathpalette\compoundreL{{#1}\over{#2}}}
\def\compoundreL#1#2{\compoundREL#1#2}
\def\compoundREL#1#2\over#3{\mathrel
   {\vcenter{\hbox{$\m@th\buildrel{#1#2}\over{#1#3}$}}}}
\makeatother
\begin{document}

\newcommand{\trS}{\mbox{$^3 \! S _1$}}
\newcommand{\trP}[1]{\mbox{$^3 \! P _{\, #1}$}}

\preprint{1.0}

\title{Heteronuclear ionizing collisions between laser-cooled 
metastable helium atoms}



\author{J.~M.~McNamara}
\author{R.~J.~W.~Stas}
\author{W.~Hogervorst}
\author{W.~Vassen}
\affiliation{Department of Physics and Astronomy, Laser Centre Vrije Universiteit, De Boelelaan 1081, 1081 HV Amsterdam, The Netherlands}


\date{\today}

\begin{abstract}
We have investigated cold ionizing heteronuclear collisions in dilute 
mixtures of metastable ($2\,\trS$) $^3\textup{He}$ and $^4\textup{He}$ 
atoms, extending our previous work on the analogous homonuclear 
collisions [R.~J.~W.~Stas \emph{et al.}, PRA \textbf{73}, 032713 
(2006)]. A simple theoretical model of such collisions enables us to 
calculate the heteronuclear ionization rate coefficient, for our 
quasi-unpolarized gas, in the absence of resonant light 
$(T=1.2\,\rm{mK})$: 
$K_{34}^{\text{(th)}}\!=\!2.4\!\times\!10^{-10}\,\rm{cm}^3/\rm{s}$. 
This calculation is supported by a measurement of $K_{34}$ using 
magneto-optically trapped mixtures containing about $1\!\times\!10^8$ 
atoms of each species, 
$K_{34}^{\text{(exp)}}\!=\!2.5(8)\!\times\!10^{-10}\,\rm{cm}^3/\rm{s}$. 
Theory and experiment show good agreement.
\end{abstract}


\maketitle

\section{Introduction}

The importance of cold collisions with regard to the dynamics of 
dilute neutral atom clouds was realized soon after the advent of 
laser-cooled and trapped atomic samples \cite{weinj03}. Since 
then, numerous investigations, both experimental and theoretical, have 
been made into the collisional properties of many different 
homonuclear 
\cite{pritd88jun,wiemc89aug,goulp92aug,aspea92dec,julip93jun,shimf93aug,shimf94nov,rolss95jan,julip95feb,walkt95aug,idot95dec,vassw06mar} 
and (later) heteronuclear 
\cite{ertmw94oct,ingum99jan,saloc99dec,vieid00jun,mosst01apr,vassw04jul2} 
systems. Collisional studies are in themselves interesting, leading to 
an in-depth understanding of the various scattering mechanisms present 
at such low kinetic energies and methods by which we can have some 
measure of control over elastic and inelastic collisions.

As opposed to collisions between thermal atoms, collisions between 
cold atoms are sensitive to the long-range part of the interatomic 
interaction potential. The de~Broglie wavelength may become comparable 
to the characteristic range of the interatomic potential, and (in the 
presence of a light field) the possibility of exciting a 
quasi-molecular state, and the subsequent decay of that state during 
a collision, becomes important. These effects lead to phenomena such as 
scattering resonances, interaction retardation, optically assisted 
collisions, photoassociation, optical shielding and the formation of 
ground-state molecules.

Optically assisted collisions lead to large losses in magneto-optical 
traps (MOTs) and are clearly dependant upon experimental parameters 
such as the intensity and detuning of the light frequencies present. A 
full theoretical treatment of these collisions is often hampered by the 
complexity of the molecular hyperfine structure \cite{weinj03}, 
thus MOTs are usually empirically optimized for their intended 
application, and comparisons with theory or other experiments are 
difficult. Collisions in the absence of a resonant light field are, on 
the other hand, a theoretically more tractable problem and their 
associated loss rate coefficients are fundamental properties of a 
given system, allowing direct comparisons with theory and between 
experiments. Homonuclear and heteronuclear collisions, "in the dark", 
between isotopes of a single element are both mediated at long-range 
by the van-der-Waals interaction ($\propto 1/\text{R}^6$) and any 
differences are due, in the main, to differing atomic structures and 
quantum statistical symmetries.

Due to the high internal energy (19.8\,eV) of metastable ($2\,\trS$) 
helium atoms (He*), the spherical symmetry of this atomic state, and 
the inverted (hyper)fine structure of the atoms; ionizing collisions 
(Penning (PI) and associative (AI)):
\begin{equation}
\begin{array}{lr}
\textup{He*}+\textup{He*}\rightarrow \textup{He}+\textup{He}^++e^- &
\mbox{(PI)},\\
\\
\nonumber \textup{He*}+\textup{He*}\rightarrow \textup{He}^{+}_2+e^- &
\mbox{(AI)},
\end{array}
\label{PenningIonisation}
\end{equation}
dominate losses in trapped samples of laser-cooled He*. This has 
provided a unique setting for the study of cold ionizing collisions in 
which the highly efficient, direct detection of collisional loss 
products using charged-particle detectors is possible. In the past, 
several experiments have made use of microchannel plate (MCP) 
detectors to measure ion production rates and investigate collisional 
losses in $^3\text{He*}$ \cite{morin99apr} and $^4\text{He*}$ 
\cite{morin99apr,nieha98jun,vassw99aug,vassw00apr}. Having realized 
the ability to trap large numbers ($>\!\!10^8$) of both isotopes 
(either individually or simultaneously) \cite{vassw04jul2}, we have 
previously reported on the isotopic differences between binary 
homonuclear collisions of $^3\text{He}$ and $^4\text{He}$ 
\cite{vassw06mar} in the absence of resonant light, resolving 
inconsistencies in prior experimental and theoretical results.

In this article we describe what we believe to be the first study of 
heteronuclear binary collisions between metastable atoms. Adapting our 
transparent theoretical model \cite{vassw06mar} slightly, we first derive 
a value for $K_{34}$ (Sec. \ref{sec:model}), the heteronuclear 
ionization rate coefficient in the absence of a resonant light field. 
This is complemented by trap loss measurements performed on a 
two-isotope magneto-optical trap (TIMOT) of $^3\textup{He*}$ and 
$^4\textup{He*}$ \cite{vassw04jul2}, from which we also extract a 
value for $K_{34}$ (Sec. \ref{sec:exp}). In Section \ref{sec:conc} we 
compare both results and briefly comment upon loss rates in the 
presence of the trapping light fields.


\section{Single channel model of the heteronuclear ionizing loss rate}
\label{sec:model}

The following model has been described in depth elsewhere in its 
successful application to the description of homonuclear ionizing 
collisions \cite{vassw06mar,stasthesis}, where it compares well with 
the more comprehensive close coupling theories developed by Venturi 
\emph{et al.} \cite{peacg99dec,whiti00may} and Leo \emph{et al.} 
\cite{babbj01sep}. Here we extend its applicability to include the 
description of heteronuclear He* collisions for which no calculations 
have previously been made. In the interests of brevity we only 
describe the models salient points and its adaption to the 
heteronuclear case.

\subsection{Ionization rate coefficients}

\begin{figure}
\includegraphics[width=0.9\columnwidth]{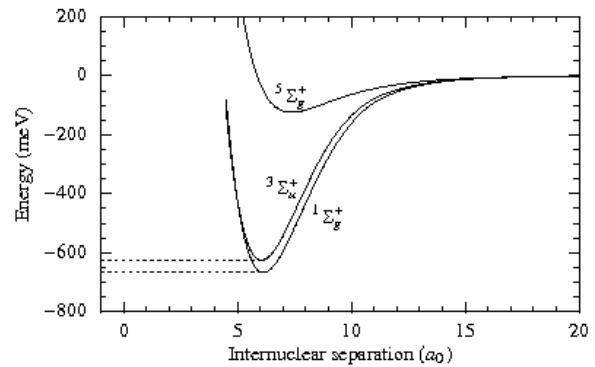}
\caption{He* potentials (labeled in Hund's case (a) notation) 
constructed as described in \cite{babbj01sep} (solid), together with 
the modifications made (dashed). Atoms reaching the region of small 
$R$ ionize; the corresponding relative particle propagates freely to 
$R=-\infty$ in our model, and ionization is accounted for by the loss 
of probability flux from the region of the potential well.}
\label{pots}
\end{figure}

\begin{figure}
\includegraphics[width=0.9\columnwidth]{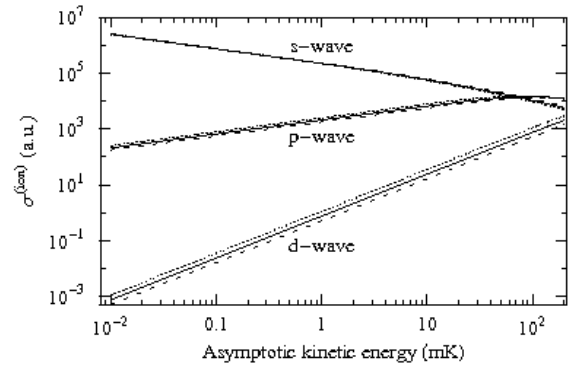}
\caption{Partial wave ionization cross sections $(S\!=\!0)$ for
\mbox{$\lsup{3}{\text{He*}}$-$\lsup{4}{\text{He*}}$} (solid lines),
\mbox{$\lsup{3}{\text{He*}}$-$\lsup{3}{\text{He*}}$} (dotted lines)
and \mbox{$\lsup{4}{\text{He*}}$-$\lsup{4}{\text{He*}}$} (dashed
lines). The familiar quantum threshold behavior
($\sigma^{\text{(ion)}}_{\ell}\!\propto\!
k^{2\ell-1}$~$(k\rightarrow 0)$) is displayed.}
\label{partialcrosssections}
\end{figure}

At mK temperatures, collisional processes are dominated by only a few partial waves, $\ell$, and the ionization cross section may be written as a sum over the partial wave contributions:
\begin{equation}
\sigma^{\text{(ion)}}=\sum_{\ell}\sigma_{\ell}^{\text{(ion)}}.
\label{PartialWaves}
\end{equation}
From a semi-classical viewpoint we may further treat the inelastic 
scattering as a two-stage process in which (at low energies) elastic 
scattering from the interaction potential $V(R)$ occurs at relatively 
large internuclear distance ($R\ge 100\, a_0$), whilst ionization 
occurs only at small internuclear distance ($R\approx 5\, a_0$) 
\cite{yenc84}. The assumption that the processes of elastic scattering 
and ionization are uncoupled allows us to factorize the probability of 
ionization occurring in a collision, and write the ionization cross 
section for collisions with total electronic spin $S$ as
\begin{equation}
\lsup{(2S+1)}{\sigma^{\text{(ion)}}}=\frac{\pi}{k^2}\sum_{\ell}(2\ell+1) \lsup{(2S+1)}{P_\ell^{\text{(tun)}}} \lsup{(2S+1)}{P^{\text{(ion)}}},
\label{CrossSection}
\end{equation}
where $k$ is the wave vector of the relative motion of the two atoms, 
$\lsup{(2S+1)}{P_{\ell}^{\text{(tun)}}}$ is the probability of the 
atoms reaching small internuclear distance (i.e., not elastically 
scattering) and \lsup{(2S+1)}{P^{\text{(ion)}}} is the probability 
of ionization occurring at short internuclear distance. In the He* 
system the quantum number $S$ is well conserved during ionization and 
the application of Wigner's spin conservation rule \cite{massh71} tells 
us that \lsup{5}{P^{\text{(ion)}}} is very small \cite{shlyg96mar}, 
while M\"{u}ller \emph{et al.} \cite{movrm91jun} report ionization 
probabilities of 0.975 for other spin states; we thus set 
$\lsup{5}{P^{\text{(ion)}}}\!=\!0$ and 
$\lsup{1}{P^{\text{(ion)}}}\!=\!\lsup{3}{P^{\text{(ion)}}}\!=\!1$ in 
Eq.~(\ref{CrossSection}). Having constructed the $\lsup{1}\Sigma^+_g$, 
$\lsup{3}\Sigma^+_u$ and $\lsup{5}\Sigma^+_g$ molecular potentials as 
described in \cite{babbj01sep} (to which we may add rotational 
barriers as required) we modify them to simulate the losses due to 
ionization (Fig.~\ref{pots}) \cite{rolss99mar}.
By numerically solving the radial wave equation and finding the 
stationary states using these potentials, we may calculate the 
incident and transmitted probability currents, $J_{\text{in}}$ and 
$J_{\text{tr}}$ respectively, the ratio of which 
($J_{\text{tr}}/J_{\text{in}}$) gives us 
$\lsup{(2S+1)}{P_{\ell}^{\text{(tun)}}}$. The calculated partial wave 
ionization cross sections are shown in 
Fig.~\ref{partialcrosssections}.

The ionization rate coefficient $K$ 
$(\textup{particle}^{-1}\,\textup{cm}^3/\textup{s})$ (determined in 
experiments) is temperature dependant, and may be written in terms of 
the ionization cross section $\sigma^{\text{(ion)}}(E)$ 
\cite{miesf89nov,rolss99mar,weinj03}:
\begin{equation}
K(T)=\int^{\infty}_{0}\sigma^{\text{(ion)}}(E)P^{\text{(MB)}}_T(v_r)\,v_r\,dv_r,
\label{IonisationRateCoefficient}
\end{equation}
where $P^{\text{(MB)}}_T(v_r)$ is the Maxwell-Boltzmann 
distribution, for a given temperature, of the relative velocities in 
the atomic sample. Similarly, we may calculate the partial wave 
ionization rate coefficients $\lsup{(2S+1)}{{\cal K}_{\ell}(T)}$ 
associated with a given \emph{molecular state} (ignoring the radial 
contributions: $|s_{1}i_{1},s_{2},S,FM_{F},\ell m_{\ell}\rangle$)~\cite{rolss99mar}.

Although the calculated values of 
\lsup{(2S+1)}{\sigma^{\text{(ion)}}_{\ell}} 
(Fig.~\ref{partialcrosssections}), and \lsup{(2S+1)}{{\cal 
K}_{\ell}(T)}, are very similar for all isotopic combinations, the 
description of the collisional process in terms of partial waves for 
each combination is very different due to the differing quantum 
statistics involved. In the case of homonuclear collisions the 
symmetrization postulate limits the number of physical scattering 
states describing a colliding pair. However, in the heteronuclear case 
there is no symmetry requirement and all partial wave contributions 
must be taken into account.

Because the energy gained during a collision is so large, the 
evolution of an atomic state 
($|s_{1}i_{1}f_{1},s_{2}j_{2},FM_{F},\ell m_{\ell}\rangle$) in the 
region where the atomic hyperfine interaction is of the same order of 
magnitude as the molecular interaction may be described to a good 
approximation as being \emph{diabatic}. Thus, to determine the 
ionization rate associated with a given \emph{scattering state} 
$K(F)$, we simply expand the atomic states onto the eigenstates of the 
short-range molecular Hamiltonian,
\begin{multline}
|s_{1}i_{1}f_{1},s_{2}j_{2},FM_{F},\ell 
m_{\ell}\rangle=\sum_{S,I}a_{SI}(F)\times\\ 
|s_{1}i_{1},s_{2},S,FM_{F},\ell m_{\ell}\rangle,
\end{multline}
determine the fraction of ionizing states $(S\!=\!0,1)$ in this 
expansion (see Table~\ref{pairstates}),
\begin{table}
\caption{Expansion coefficients $a_{SI}(F)\!=\!\langle s_1 i_1, s_2 ,S, 
F M_F, \ell m_\ell | s_1 i_1 f_1,s_2 j_2, F M_F, \ell m_\ell \rangle$. 
The scattering states $|s_1 i_1 f_1,s_2 j_2, F M_F, \ell m_\ell 
\rangle$ are indicated by their values of $F$, while the molecular 
states $|s_1 i_1, s_2,S, F M_F, \ell m_\ell \rangle$ are given in 
Hund's case (a) notation, $^{2S+1}\Sigma_{\text{g/u}}^+$.}
\begin{ruledtabular}
\begin{tabular}{lccc}
F&$^{1}\Sigma_{\text{g}}^+$ &$^{3}\Sigma_{\text{u}}^+$
&$^{5}\Sigma_{\text{g}}^+$ \rule[-2mm]{0mm}{6mm}\\ \hline
1/2&$\sqrt{2/3}$&$-\sqrt{1/3}$&\rule[0mm]{0mm}{5mm}\\
3/2&&$\sqrt{5/6}$&$-\sqrt{1/6}$\rule[0mm]{0mm}{5mm}\\
5/2&&&$1$\rule[0mm]{0mm}{5mm}\\
\end{tabular}
\end{ruledtabular}
\label{pairstates}
\end{table}
and sum over the contributing partial waves:
\begin{equation}
K(F)=\sum_{\ell}\sum_{S,I}|a_{SI}(F)|^2\times \lsup{(2S+1)}{{\cal 
K}_{\ell}}.
\end{equation}
Using the partial wave ionization rates obtained we may then derive 
the total ionization rate coefficient for an unpolarized sample, i.e., 
the rate coefficient for which the magnetic substates of the atoms in 
our sample are evenly populated:
\begin{equation}
K^{\text{(unpol)}}=\frac{1}{(2f_1+1)}\frac{1}{(2f_2+1)}\sum_{F}\sum_{M_F}K(F),
\label{ThRateCoeff}
\end{equation}
which in the heteronuclear case gives: 
\begin{equation}
K^{\text{(unpol)}}_{34}\approx\frac{1}{12}\left[\frac{4}{3}(\lsup{1}{{\cal 
K}}_0+\lsup{1}{{\cal K}}_1)+{4}(\lsup{3}{{\cal K}}_0+\lsup{3}{{\cal K}}_1)\right].
\label{ApproxThRateCoeff}
\end{equation}
\begin{figure}
\includegraphics[width=0.9\columnwidth]{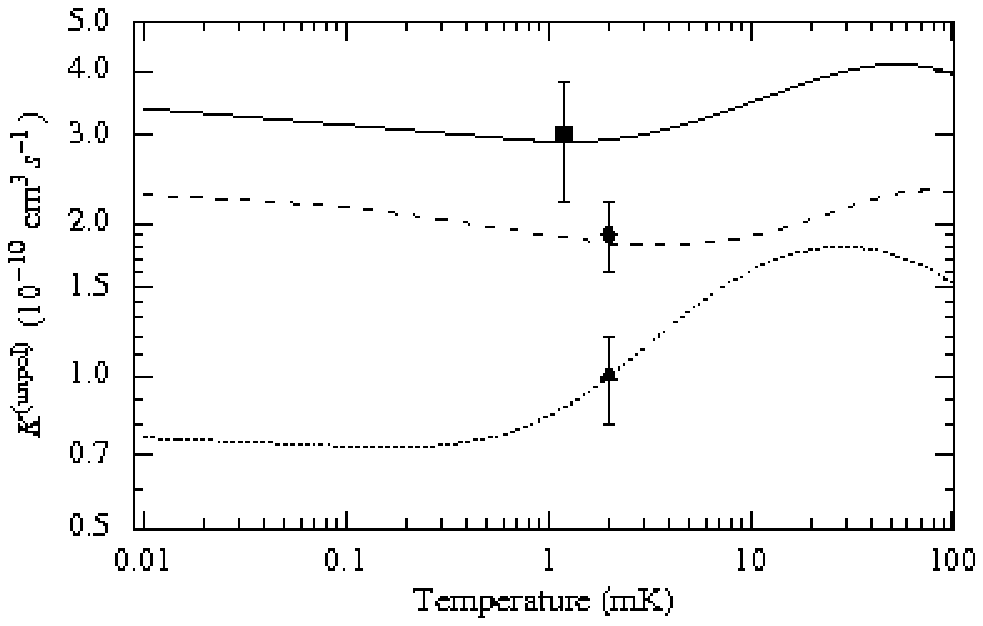}
\caption{Theoretical unpolarized loss rate coefficient curves:
\mbox{$\lsup{3}{\text{He}}\,\text{\textendash}\,\lsup{3}{\text{He}}$} 
(dashed), 
\mbox{$\lsup{4}{\text{He}}\,\text{\textendash}\,\lsup{4}{\text{He}}$} 
(dotted) and, 
\mbox{$\lsup{3}{\text{He}}\,\text{\textendash}\,\lsup{4}{\text{He}}$} 
(solid), together with our experimental data points and their error 
bars: 
\mbox{$\lsup{3}{\text{He}}\,\text{\textendash}\,\lsup{3}{\text{He}}$} 
(diamond), \mbox{$\lsup{4}{\text{He}}\,\text{\textendash}\,\lsup{4}{\text{He}}$}
(triangle) and 
\mbox{$\lsup{3}{\text{He}}\,\text{\textendash}\,\lsup{4}{\text{He}}$}
(square). For the purposes of this figure, the experimental points have 
been corrected for the inhomogeneous distribution over the magnetic 
substates found in our MOT's (see Sec.~\ref{trappedsamples}).}
\label{ratecoefficients}
\end{figure}
The energy dependant unpolarized ionization rate coefficients are 
shown in Fig.~\ref{ratecoefficients} (including those rates previously 
calculated \cite{vassw06mar} for $^3\textup{He*}$ and $^4\textup{He*}$) 
and for a temperature of $T\!=\!1\,\text{mK}$ we obtain
\begin{equation}
K^{\text{(unpol)}}_{34}=2.9\times 10^{-10}~\text{cm}^3/\text{s}.
\end{equation}

\subsection{Ionization rate coefficient of trapped samples}
\label{trappedsamples}

Optical pumping processes in a MOT cause the distribution over 
magnetic substates $P_{m}$ (where $m$ is the azimuthal quantum number) 
of the trapped atoms to differ from the uniform (unpolarized) 
distribution assumed above. This is important as the contribution of 
each collision channel to the ionizing losses depends upon $P_{m}$, 
and can be accounted for in our theoretical model by using the density 
operator \cite{cohec77}
\begin{equation}
\rho(\bm{r}) = \sum_m \sum_{n\leq m} P_m(\bm{r}) \, P_n(\bm{r}) \, |m, n \rangle \langle m,n |
\label{eq:densityoperator}
\end{equation}
to describe a statistical mixture of magnetic substate pairs 
$|m,n\rangle$, where $m$ and $n$ are the azimuthal quantum numbers of 
the colliding atoms. The ionization rate coefficient of the mixture 
can then be written as
\begin{equation}
K_{34} = \frac{1}{N_{3}N_{4}} \iiint \Bigl( \sum_{\ell} \bigl(\lsup{1}{b\onemu} 
\: \lsup{1}{\cal K}_\ell + \lsup{3}{b\onemu}\: \lsup{3}{\cal K}_\ell 
\bigr)\Bigr)n_{3}(\bm{r})n_{4}(\bm{r}) \, \text{d}^3r,
\label{eq:Karb}
\end{equation}
where $N_{3}$ and $N_{4}$ are the number of trapped atoms in each 
component of the mixture, the coefficients $\lsup{(2S+1)}{{b\onemu}}$ 
are the sums of the expectation values of the density operator for all 
ionizing molecular states with total spin $S$, and, $n_{3}(\bm{r})$ and 
$n_{4}(\bm{r})$ are the density distributions of each component in the 
sample. Explicit expressions for the coefficients 
$\lsup{(2S+1)}{{b\onemu}}$ are given in Table~\ref{coefficients}.
\begin{table*}
\caption{$^{3}\text{He*}$-$^{4}\text{He*}$
$\lsup{(2S+1)}{{b\onemu}}$ coefficients from
Eq.~(\ref{eq:Karb}). The coefficients are the expectation values of 
the density operator, Eq.~(\ref{eq:densityoperator}), for all ionizing 
molecular states of given $S$ and parity.}
\begin{ruledtabular}
\begin{tabular}{rc}
$^{(2S+1)}b$&$^{3}\text{He*}$-$^{4}\text{He*}$
\rule[-2mm]{0mm}{6mm}\\ \hline
$^{1}b$&$(1/3)(P_{-3/2}P_{1}+P_{3/2}P_{-1})+(2/9)(P_{-1/2}P_{0}+P_{1/2}P_{0})+(1/9)(P_{-1/2}P_{1}+P_{1/2}P_{-1})$\rule[0mm]{0mm}{5mm}\\
$^{3}b$&$(1/2)(P_{-3/2}P_{1}+P_{-3/2}P_{0}+P_{-1/2}P_{1}+P_{1/2}P_{-1}+P_{3/2}P_{0}+P_{3/2}P_{-1})$\rule[0mm]{0mm}{5mm}\\
&$+(1/3)(P_{-1/2}P_{-1}+P_{1/2}P_{1})+(1/6)(P_{-1/2}P_{0}+P_{1/2}P_{0})$\\
\end{tabular}
\end{ruledtabular}
\label{coefficients}
\end{table*}
One may easily check that we recover Eq.~(\ref{ApproxThRateCoeff}) 
from Eq.~(\ref{eq:Karb}) by substituting the values 
$P_{-3/2}\!=\!P_{-1/2}\!=\!P_{1/2}\!=\!P_{3/2}\!=\!\frac{1}{4}$ and 
$P_{-1}\!=\!P_{0}\!=\!P_{1}\!=\!\frac{1}{3}$ into the expressions for 
the coefficients $\lsup{(2S+1)}{b}$ (Table~\ref{coefficients}) and 
evaluating Eq.~(\ref{eq:Karb}).

In order to later make a comparison between theory and experiment, we 
determine the distribution $P_{m}(\bm{r})$ by obtaining the 
steady-state solution of a rate equation model describing the optical 
pumping in our MOT \cite{vassw00apr}. At a temperature of 
$T\!=\!1.2\,\text{mK}$ the resulting value of the theoretical 
ionization rate coefficient is
\begin{equation}
K^{\text{(th)}}_{34}=2.4\times 10^{-10}~\text{cm}^3/\text{s}
\end{equation}

\section{Experimental Setup}

\begin{figure}
\includegraphics[width=0.9\columnwidth]{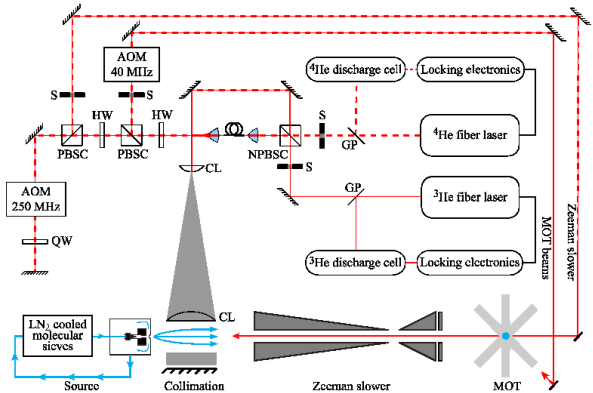}
\caption{A schematic overview of the TIMOT experimental apparatus used 
in these experiments. Component labels are: AOM, acousto-optic 
modulator; QW, quater-wave plate; S, shutter; PBSC, polarizing beam 
splitter cube; HW, half-wave plate; CL, cylindrical lens; and NPBSC, 
non-polarizing beamsplitter cube. For clarity, all spherical lenses 
and excess mirrors have been omitted. }
\label{fig:setup}
\end{figure}

We investigate cold ionizing collisions in a setup (see 
Fig.~\ref{fig:setup}) capable of trapping large numbers ($\gsim10^8$) 
of both $^3$He* and $^4$He* atoms simultaneously in a TIMOT 
\cite{vassw06mar}. A collimated and Zeeman slowed He* beam is used to 
load our TIMOT which is housed inside a stainless steel, ultra-high 
vacuum chamber. The beam is produced by a liquid nitrogen (LN$_2$) 
cooled dc discharge source supplied with an isotopically enriched 
($\approx\! 50/50$) gaseous mixture of $^3\textup{He}$ and 
$^4\textup{He}$ held in a helium tight reservoir. During operation the 
reservoir is connected such that all helium not entering the 
collimation section (the vast majority of it) is pumped back into the 
reservoir and recycled, conserving our supply of the relatively 
expensive $^3$He gas. Two LN$_2$ cooled molecular sieves, ensuring a 
pure supply of helium to the source, are also contained within this 
reservoir; the first of these is a type 13X molecular sieve, whilst 
the second is type 4A (both are sodium zeolites having pore sizes of 
$10\,\text{\AA}$ and $4\,\text{\AA}$ respectively). We then make use 
of the curved wavefront technique to collimate our atomic beam in two 
dimensions \cite{vassw96jan} before it enters the Zeeman slower. Due 
to its lighter mass $^3$He atoms emerge from the source with a greater 
mean velocity than $^4$He atoms and in order to achieve a large flux 
of both $^3$He* and $^4$He* atoms we have increased the capture 
velocity of our Zeeman slower \cite{stasthesis,vassw04jul2}. Our 
ultra-high vacuum chamber maintains an operational pressure of $7 
\times 10^{-10}\,\text{mbar}$ (with a partial presure of $6.5\times 
10^{-10}\,\text{mbar}$, ground state He atoms from the atomic beam are 
the major contribution to this) and is based upon the design of our 
next-generation BEC chamber \cite{vassw06mar2}. Two coils in an 
anti-Helmholtz configuration produce the quadrupole magnetic field 
($dB/dz\!=\!0.35\,\text{T/m}$) for the TIMOT; these are placed in 
water cooled buckets outside the vacuum and are brought close to the 
trapping region by placing them inside reentrant glass windows 
situated on either side of the chamber.

\begin{figure}
\begin{center}
\includegraphics[width=0.75\columnwidth]{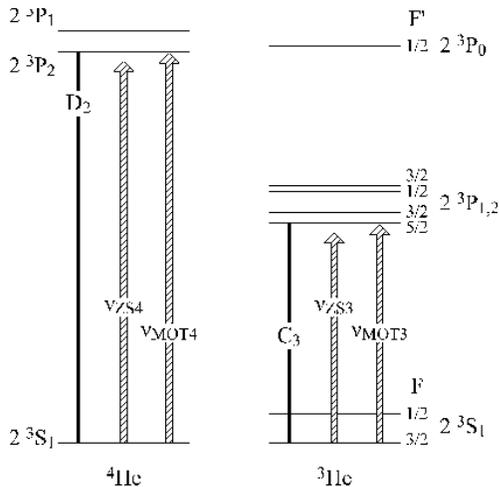}
\caption{Level scheme for the ground and first excited states in
$^4\textup{He}^*$ and $^3\textup{He}^*$.}
\label{transitionspic}
\end{center}
\end{figure}

Both helium isotopes are collimated, slowed, and confined in the TIMOT 
using 1083~nm light nearly resonant with their $2\,\trS \rightarrow 
2\,\trP{2}$ optical transitions (see Fig.~\ref{transitionspic}) 
(natural linewidth $\Gamma/2\pi\!=\!1.62\,\text{MHz}$ and saturation 
intensity $I_{\text{sat}}\!=\!0.166\,\text{mW\,cm}^{-2}$, for the cycling 
transition).
As the isotope shift for this transition is $\approx\!34\,\text{GHz}$ 
the bichromatic beams are produced by overlapping the output from two 
ytterbium-doped fiber lasers (IPG Photonics) on a non-polarizing 50/50 
beam splitter. One beam is sent to the collimation section, whilst the 
second is coupled into a single mode polarization maintaining fiber to 
ensure a perfect overlap of the two frequency components, before being 
split into the Zeeman slowing beam and the trapping beams. Each fiber 
laser is locked to the respective cooling transition using saturated 
absorption spectroscopy in an rf-discharge cell; acousto-optic 
modulators are then used to generate the slowing and trapping 
frequencies which are detuned by -500\,MHz and -40\,MHz respectively 
(see Fig.~\ref{transitionspic}). The slowing beam is focused on the 
source and has a 1/$e^2$ intensity width of 2.2\,cm at the position of 
the trapped cloud, with each frequency component having a peak 
intensity of $I_{\text{peak}}\!=\!9\,\text{mW\,cm}^{-2}$ 
($I_{\text{peak}}/I_{\text{sat}}\!\approx\!54$), while the trapping 
beam is split into six independent Gaussian beams with 1/$e^2$ 
intensity widths of 1.8\,cm and total peak intensity 
$I_{\text{peak}}\!=\!57\,\text{mW\,cm}^{-2}$ 
($I_{\text{peak}}/I_{\text{sat}}\!\approx\!335$). The outputs of two 
diode lasers (linewidths $<500\,\text{kHz}$), each locked using 
saturated absorption spectroscopy to the helium $2\,\trS \rightarrow 
2\,\trP{2}$ resonance of one of the isotopes, are overlapped on a 
second non-polarizing beam-splitter cube and coupled into a single 
mode polarization maintaining fiber. This system delivers two, weak, 
linearly polarized probe beams ($\Delta\!=\!0$, 
$I\!=\!0.05\,I_{\text{sat}}$, with 
$I_{\text{sat}}\!=\!0.27\,\text{mW\,cm}^{-2}$ assuming equal 
population of all magnetic sublevels of the $2\,\trS$ state in the 
trap) to the chamber for the absorption imaging of each component of 
the trapped cloud. Absorption images are recorded using an 
IR-sensitive charge-coupled device (CCD) camera \cite{tolthesis}. 
Unfortunately the imaging of both trapped components cannot be carried 
out in a single experimental cycle; sequential runs must be made, 
imaging first one component and then the other.

Trapped clouds are further monitored by two microchannel plate (MCP) 
detectors mounted inside the chamber. Operated at a voltage of 
-1.5\,kV and positioned 11\,cm from the center of the trap, the MCP's 
are used to independently monitor the ions and $\text{He}^*$ atoms 
escaping or released from the trap. With an exposed front plate held 
at negative high voltage, one MCP mounted above the trap center 
attracts all positive ions produced during ionizing collisions (Eq. 
(1)) in the trap. The second MCP is shielded by a grounded grid, 
mounted below the trap center, and detects only $\text{He}^*$ atoms.

The shielded MCP is used to perform time-of-flight (TOF) measurements 
from which we determine the temperature of the trapped atoms. An 
absorpton image, in combination with the measured temperature, then 
allows us to determine the density distribution, size, and absolute 
atom number of the sample. We then use the unshielded MCP to measure 
the instantaneous ionization rate in the trapped sample, which, in 
combination with the information obtained from the absorption images, 
allows us to determine trap loss and ionization rates in the sample. 
Our typical TIMOT parameters and measurements have been reported 
previously \cite{vassw04jul2,vassw06mar} and are given in 
Table~\ref{oldrestable}. The lower temperatures realized in the 
present experiments are due to the resolution of a power imbalance in 
two of the trapping laser beams.
\begin{table}
\caption{Typical experimental values of the \lsup{3}{\text{He*}} and 
\lsup{4}{\text{He*}} components in our TIMOT (error bars correspond to 
1 standard deviation).\\}
\label{oldrestable}
\begin{ruledtabular}
\begin{tabular}{lcc}
&$^3\textup{He*}$ &$^4\textup{He*}$\rule[-2mm]{0mm}{6mm}\\ \hline
Temperature \textit{T} (mK) &1.2(1) &1.2(1)\rule[0mm]{0mm}{5mm}\\
Number of atoms \textit{N} &$1.0(3) \times 10^8$ &$1.3(3) \times
10^8$\rule[0mm]{0mm}{5mm}\\
Central density $n_0$ (cm$^{-3}$) &$0.5(1) \times 10^9$ &$1.0(2) \times 10^9$\rule[0mm]{0mm}{5mm}\\
Axial radius $\sigma_p$ (cm) &0.28(4) &0.25(3)\rule[0mm]{0mm}{5mm}\\
Radial radius $\sigma_z$ (cm) &0.16(2) &0.14(1)\rule[0mm]{0mm}{5mm}\\
\end{tabular}
\end{ruledtabular}
\end{table}

\section{Determination of the Heteronuclear Loss Rate}
\label{sec:exp}

The time evolution of the total number of atoms trapped in our TIMOT, 
$N\!=\!N_3\!+\!N_4$, may be described by the following phenomenological 
equation \cite{pritd88jun}:
\begin{widetext}
\begin{equation}
\frac{dN}{dt}=L_3-\alpha_3 N_3(t)-\beta_{33}\int\!\!\!\int\!\!\!\int 
n_3^2(\textbf{r},t)\,d^3\textbf{r}+L_4-\alpha_4N_4(t)-\beta_{44}\int\!\!\!\int\!\!\!\int 
n_4^2(\textbf{r},t)\,d^3\textbf{r}-\beta_{34}\int\!\!\!\int\!\!\!\int n_3(\textbf{r},t)n_4(\textbf{r},t)\,d^3\textbf{r},
\label{NumAtoms}
\end{equation}
\end{widetext}
where $t$ denotes time, $L$ is the rate at which atoms are loaded 
into the TIMOT, $\alpha$ is the linear loss rate coefficient describing 
collisions between trapped He* atoms and background gases, $\beta$ is 
the \emph{loss rate} coefficient resulting from binary collisions 
between trapped He* atoms, and $n$ is the cloud density. Subscripts 
denote whether a given parameter pertains to either of the $^3$He* or 
$^4$He* components of the mixture.

It should be noted at this point that the "density limited" regime, 
often mentioned with regard to the alkali systems, is not a feature of 
the metastable noble gas systems; in the latter the density is not 
limited by radiation trapping within the cloud, but by ionizing 
collisional losses. This is born out in Eq.~(\ref{NumAtoms}) by the 
time dependance of the atom density distributions and hence our 
inability to make the simplifying constant density approximation in 
the following experiments.

Both linear and quadratic losses in Eq.~(\ref{NumAtoms}) are due to a 
number of different mechanisms, and may be subdivided into either 
ionizing or non-ionizing categories. The ion production rate may then 
be expressed in a manner analogous to Eq.~(\ref{NumAtoms}):
\begin{widetext}
\begin{equation}
\frac{dN_{\text{ion}}}{dt}=\epsilon_a\alpha_3
N_3(t)+\epsilon_bK_{33}\int\!\!\!\int\!\!\!\int
n_3^2(\textbf{r},t)\,d^3\textbf{r}+\epsilon_c\alpha_4N_4(t)+\epsilon_dK_{44}\int\!\!\!\int\!\!\!\int
n_4^2(\textbf{r},t)\,d^3\textbf{r}+\epsilon_eK_{34}\int\!\!\!\int\!\!\!\int n_3(\textbf{r},t)n_4(\textbf{r},t)\,d^3\textbf{r},
\label{NumIons}
\end{equation}
\end{widetext}
where $\epsilon_a, \epsilon_b, \epsilon_c, \epsilon_d$ and 
$\epsilon_e$ are the weights of the various ionization mechanisms (and 
may include a factor to account for a less than unity detection 
efficiency), and the \emph{collision} rate coefficient $K$ has been 
introduced. The \emph{loss} and \emph{collision} rate coefficients 
are related by the equation $\beta = 2K$, and the appearance of $K$, 
instead of $\beta$ in Eq.~\ref{NumIons} expresses the fact that during 
each ionizing collision, one ion is produced, but two atoms are lost 
from the trap. From an analysis of the trap loss mechanisms 
\cite{vassw06mar,stasthesis} it can be seen that to a good 
approximation $\epsilon_a\!=\!\epsilon_c\!=\!0$, whilst 
$\epsilon_b\!=\!\epsilon_d\!=\!\epsilon_e$. The current signal 
measured by the MCP is proportional to the ionization rate, 
Eq.~(\ref{NumIons}); and for gaussian spatial density distributions 
(centered with respect to each other), the voltage measured by the 
oscilloscope may then be written as
\begin{widetext}
\begin{equation}
\phi_{TIMOT} 
(t)=eR_{\text{eff}}\left[K_{33}\,n^2_{03}(t)(\pi\sigma_3^2)^{\frac{3}{2}}+K_{44}\,n^2_{04}(t)(\pi\sigma_4^2)^{\frac{3}{2}}+K_{34}\,n_{03}(t)n_{04}(t)\left[\frac{2\pi\sigma^2_3\sigma^2_4}{\sigma_3^2+\sigma_4^2}\right]^\frac{3}{2}\right]+\phi_{\text{bgr}},
\label{MCPVolts}
\end{equation}
\end{widetext}
where $n_0$ is the central density, $\sigma$ is the mean rms radius 
(of a given cloud component), $e$ is the electron charge, and 
$R_{\text{eff}}$ is an effective resistance. 

\begin{figure}
\includegraphics[width=0.9\columnwidth]{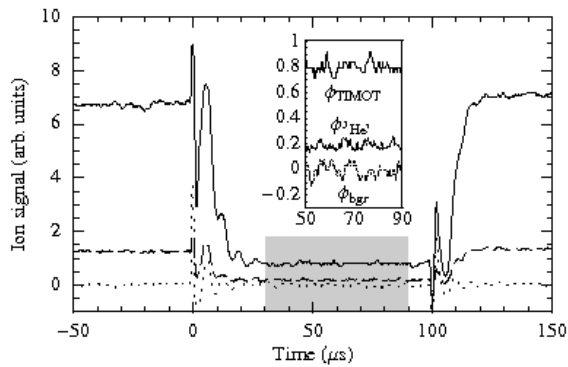}
\caption{Averaged ion signals measured during the experiments. The 
full line is the signal recorded whilst performing the experiment 
with a TIMOT, the dashed line is the corresponding curve for a 
\lsup{3}{\text{He*}} MOT (the analogous signal for a 
\lsup{4}{\text{He*}} MOT can also be obtained), and the dotted line is 
the background signal. The ion rate measured is then averaged over the 
$60\,\mu s$ interval indicated by the shaded region in order to obtain 
($\phi_{\text{TIMOT}}\!-\!\phi_{\text{bgr}}$) and 
($\phi_{^3\text{He*/}^4\text{He*}}\!-\!\phi_{\text{bgr}}$). }
\label{dutycyclepic}
\end{figure}

The experiment is based around the ability of an MCP detector to 
measure the ions produced in our metastable isotopic mixture with very 
high efficiency, and employs a method first used by Bardou \emph{et 
al.}~\cite{aspea92dec} to determine the ionization rate in the absence 
of light. To measure the rate in the dark we perform an experiment in 
which we load the TIMOT, switch off the Zeeman slower beam and all MOT 
beams using the frequency detuning AOM's (the quadrupole field remains 
on) for $100\,\mu\text{s}$, before switching the slowing and trapping 
light back on again. This on/off cycle is easily repeated many times 
while we monitor the ion signal and average it (see 
Fig.~\ref{dutycyclepic}). The switch-off time is short compared to the 
dynamics of the expanding cloud (we see no variation in the ion signal 
during the switch off period) and we switch the light on long enough 
(200\,ms) to recapture the cloud and allow it to equilibrate.

Equation~(\ref{MCPVolts}) describes the ion production rate in the 
dark of our TIMOT; a similar equation describes the ion production 
rate of a single-isotope MOT, $\phi_{MOT}$. By combining the equations 
for $\phi_{TIMOT}$ and $\phi_{MOT}$ with a measurement of the ratio 
$r=(\phi_{TIMOT}-\phi_{bgr})/(\phi_{MOT}-\phi_{bgr})$ (where the time 
dependance of the measured signals has been omitted because of the 
short duration of the switch-off period), we can derive an expression 
for $K_{34}$. It has been verified that under our experimental 
conditions the MCP signal varies linearly with the ion production 
rate, and all cloud densities and radii may be  derived from 
absorption images, while we have previously measured $K_{33}$ and 
$K_{44}$~\cite{vassw06mar}. As the trap parameters, and therefore the 
distribution over the magnetic substates of the atoms, have changed 
since the experiments described in Ref.~\cite{vassw06mar} were 
performed, we have used the theory described in 
Sec.~\ref{trappedsamples} to correct the measured values of $K_{33}$ 
and $K_{44}$ for these effects, yielding: $K_{33}\!=\!1.6(3)\times 
10^{-10}\,\text{cm}^3/\text{s}$ and $K_{44}\!=\!6.5(2)\times 
10^{-11}\,\text{cm}^3/\text{s}$.



\section{Results and discussion}
\label{sec:conc}

Performing the experiments described in the previous section, we obtain 
the result: $K_{34}^{(exp)}\!=\!2.5(8)\times 
10^{-10}\,\text{cm}^3/\text{s}$, which compares well with the 
theoretical prediction, 
$K^{\text{(th)}}_{34}\!=\!2.4\times 10^{-10}\,\text{cm}^3/\text{s}$, 
obtained after undertaking the calculation described in 
Sec.~\ref{sec:model}. For future comparison with other theoretical 
models, the equivalent theoretical loss-rate coefficient for an 
unpolarized gas mixture at a temperature of $T\!=\!1\,\rm{mK}$ is 
$K_{34}^{(unpol)}\!=\!2.9\times 10^{-10}\,\text{cm}^3/\text{s}$ (see 
Fig.~\ref{ratecoefficients}).These numbers describe the \emph{total} 
heteronuclear loss rate in the absence of light; the only other 
potential loss process would be hyperfine state changing collisions, 
however, due to the inverted nature of the hyperfine structure in 
$^3\textup{He*}$ the atoms occupy the lowest hyperfine level of the 
$2\,\trS$ multiplet and so cannot relax to a lower level, while the 
endothermic collision necessary to reach the $\text{F}\!=\!1/2$ state 
would require $\approx\! 200\,\text{mK}$ to be provided by the trap, 
200 times more than the typical collision energy in the TIMOT. The 
error in the experimental value is mainly determined by errors in the 
absorption imaging, which we find to be $\approx\! 30\%$, and by the 
error bars reported on our previous measurements of the homonuclear 
loss rates~\cite{vassw06mar}. 

With regard to optically assisted collisions, we note that the excited 
state potentials in the case of heteronuclear collisions are governed 
at long-range by the van-der-Waals interaction 
($\propto\!1/\text{R}^6$), having a much shorter range than that of 
the resonant dipole interaction ($\propto\!1/\text{R}^3$) dominant in 
the homonuclear case. In our TIMOT the laser beams are, in contrast to 
most experiments performed on heteronuclear collisions, far-detuned 
(25 natural linewidths) from resonance and the atomic excited state 
population in the trap is therefore negligible. We can only excite a 
molecular state if the correct light frequency is present and the 
atoms have reached the Condon point for the transition. As all light 
frequencies in our TIMOT are far detuned from any heteronuclear 
transition we expect no contribution to the trap loss rate from 
optically assisted heteronuclear collisions.

To summarize, we have measured the heteronuclear loss rate 
coefficient $K_{34}^{(exp)}$ in the absence of light for a trapped 
mixture of $^3\textup{He*}$ and $^4\textup{He*}$ atoms at 
$T\!=\!1\,\text{mK}$. The measured value of $K_{34}^{(exp)}$ compares 
very well with the value predicted by our single channel model of 
ionizing collisions in the He* system. Recently, both helium isotopes 
were magnetically trapped in the multi-partial wave regime using 
buffer-gas cooling, and a deep magnetic trap~\cite{doylj05dec}. The 
probable observation of Penning ionization under these conditions has 
been reported~\cite{nguyenthesis}, and  it would be interesting to 
extend our theory into the multi-partial wave regime and to high 
magnetic field values. With the production of quantum degenerate 
mixtures of \lsup{3}{\text{He*}} and 
\lsup{4}{\text{He*}}~\cite{mcnaj06aug}, we also have the possibility 
of investigating both homonuclear and heteronuclear He* collisions at 
ultracold temperatures in greater detail. An experiment of interest in 
this area would be the implementation of an optical dipole trap in 
which it would be possible to prepare ultracold samples in well 
defined magnetic substates. In particular, it would be possible to 
prepare trapped ultracold samples of \lsup{3}{\text{He*}} 
(\lsup{4}{\text{He*}}) in the $m_F\!=\!-3/2$ ($m_J\!=\!-1$) states, 
for which (as in the magnetically trapped $m_F\!=\!+3/2$ 
($m_J\!=\!+1$) states used in the production of ultracold He* gases) 
Penning ionization should be suppressed.

\begin{acknowledgments}
We thank Jacques Bouma for technical support. This work was supported 
by the ``Cold Atoms'' program of the Dutch Foundation for Fundamental 
Research on Matter (FOM), the Space Research Organization Netherlands 
(SRON), Grant No. MG-051, and the European Union, Grant No. 
HPRN-CT-2000-00125.
\end{acknowledgments}


\end{document}